\documentstyle[aps,prd,eqsecnum,epsf]{revtex}
\newcommand{\nnb}{\nonumber}
\newcommand{\V}{\nabla\!}
\newcommand{\ppp}{\partial}

\begin{document}
\thispagestyle{empty}

%
%
\leftline{\epsfbox{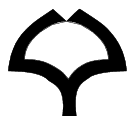}}
\vspace{-10.0mm}
{\baselineskip-4pt
\font\yitp=cmmib10 scaled\magstep2
\font\elevenmib=cmmib10 scaled\magstep1  \skewchar\elevenmib='177
\leftline{\baselineskip20pt
\hspace{12mm} 
\vbox to0pt
   { {\yitp\hbox{Osaka \hspace{1.5mm} University} }
     {\large\sl\hbox{{Theoretical Astrophysics}} }\vss}}

%
%
{\baselineskip0pt
\rightline{\large\baselineskip14pt\rm\vbox 
        to20pt{\hbox{\today}
               \hbox{OU-TAP-150}
               \hbox{UTAP-379}
\vss}}
}
\vskip15mm

\begin{center}
{\Large\bf Radion on the de Sitter brane}
\end{center}

\begin{center}
{\large Uchida Gen$^{1,2}$ and Misao Sasaki$^1$ }
\vskip 3mm
${}^1$\sl{Department of Earth and Space Science, Graduate School of Science,\\
Osaka University, Toyonaka 560-0043, Japan}
\vskip 3mm
${}^2$\sl{Department of Physics, School of Science,\\
University of Tokyo, Tokyo 113-0033, Japan}
\end{center}


\begin{abstract} 
The radion on the de Sitter brane is investigated at the linear
perturbation level, using the covariant curvature tensor formalism
developed by Shiromizu, Maeda and Sasaki\cite{SMS}.
It is found that if there is
only one de Sitter brane with positive tension, there is no radion and
thus the ordinary Einstein gravity is recovered on the brane other than
corrections due to the massive Kaluza-Klein modes.
As a by-product of using the covariant curvature tensor formalism,
it is immediately seen that the cosmological scalar, vector and tensor
type perturbations all have the same Kaluza-Klein spectrum.
On the other hand, if there are two branes with positive and
negative tensions, the gravity on each brane takes corrections from the
radion mode in addition to the Kaluza-Klein modes and the radion is
found to have a negative mass-squared proportional to the curvature
of the de Sitter brane, in contrast to the flat brane case in which
the radion mass vanishes and degenerates with the $4$-dimensional
graviton modes.
To relate our result with the metric perturbation approach,
we derive the second order action for the brane displacement.
We find that the radion identified in our approach indeed corresponds to
the relative displacement of the branes in the Randall-Sundrum gauge
and describes the scalar curvature perturbations of the branes
in the gaussian normal coordinates around the branes. 
Implications to the inflationary brane universe are briefly discussed.
\end{abstract}


\section{Introduction}

Based on the idea of a brane-world suggested from string
theory\cite{Antoniadis},
Randall and Sundrum proposed an interesting scenario that we may live
on a 3-brane embedded in the 5-dimensional anti-de Sitter
space\cite{RS1,RS2}. One of the attractive features of their scenario
is that gravity on the brane may be confined within a short distance
from the brane to be effectively 4-dimensional even for an infinitely
large extra-dimension\cite{RS2,GT}.
 This occurs on a brane with positive tension,
and it is because the AdS bulk on both sides of the brane
shrinks exponentially as one goes away from the brane.

In their comprehensive paper\cite{GT}, Garriga and Tanaka has shown
that if there is only one flat brane with positive tension,
the zero-mode truncation, which means to discard all
the excitation modes except for the one with the lowest (zero)
eigenvalue, leads to 4-dimensional (linearized) Einstein gravity 
on the brane. 
 On the other hand, if there are two flat branes of positive and
negative tensions, the zero-mode truncation leads to a Brans-Dicke type
theory with positive and negative Brans-Dicke parameters, respectively,
on the branes with positive and negative tensions. The difference can be
understood by considering the role of the so-called
radion mode, or the unique non-trivial scalar mode of gravitation in the
vacuum bulk spacetime.
(Hereafter, we use the term, scalar, in the sense of the 4-dimensions
unless otherwise noted.)
 The radion mode corresponds to displacement of the
brane location\cite{GT,CGR}. If only a single flat brane is present, 
it is impossible to know locally if the brane is displaced because
of the general covariance. Hence, the radion plays no role in the
gravity on the brane. However, if two flat branes are present, the
relative motion of the branes, or the relative displacement of the
branes is physical. Therefore the radion mode is expected to affect
the gravity on the brane. 
Thus if we live on a single positive-tension
brane, the effect of the extra-dimension appears only
through the non-zero-mode excitations, or the so-called
Kaluza-Klein modes. 

However, the above interpretation is given
to the case of flat Minkowski branes. It is therefore interesting
to ask if the same argument applies to the case of cosmological
brane models, particularly when the universe is spatially closed. 
It was suggested that a spatially closed,
inflating brane-universe may be created from nothing\cite{GS}.
In this model, the universe is described by the de Sitter brane embedded
in the 5-dimensional anti-de Sitter space.
Since this brane-universe is spatially closed, the curvature radius
of the universe becomes indeed the 5-dimensional radius of the brane.
Then, one may expect that the radion, which describes fluctuations
of the brane location, becomes physical even in the case of a single
brane, and contributes to gravity on the brane.

In this paper, we carefully investigate perturbations of the de Sitter
brane-world and clarify this issue of the cosmological radion.
Cosmological brane perturbations have been studied
by various authors with metric-based
approaches\cite{GS,Mukohyama,Kodama,HHR,Carsten,Koyama,Langlois,%
MWBH,LMW,bmw,cll,DDK,NeroSachs}.
We adopt the covariant curvature tensor
formalism developed by Shiromizu, Maeda and Sasaki\cite{SMS}. 
A covariant approach to cosmological brane perturbations
based on this formalism has been developed by Maartens\cite{Roy}.
In this formalism, the effective gravitational equation is written in
the form of the 4-dimensional Einstein equations with a couple of
modifications on the right-hand-side.
Namely, there appear a tensor quadratic in the matter energy-momentum
tensor and an ``electric'' projection of the 5-dimensional Weyl tensor.
Since the former is locally described by
the 4-dimensional energy-momentum tensor, the genuine 5-dimensional
gravitational effect is contained in the projected Weyl tensor.
We denote this tensor by $E_{\mu\nu}$ 
(see Eq.~(\ref{Edefined}) for its definition).

A noble feature of this formalism is that the perturbations
that describe the 4-dimensional gravitational degrees of freedom
are already contained in the effective
4-dimensional Einstein equations and the perturbations that additionally
arise from the existence of the 5-dimensional bulk
are completely described by the projected Weyl tensor
mentioned above. Another advantage of the formalism is that
since the 5-dimensional Weyl tensor vanishes on the anti-de Sitter
background, $E_{\mu\nu}$ is a physical, gauge-invariant quantity
by itself.

In this paper we neglect the tensor quadratic in the matter
energy-momentum and investigate the behavior of the projected Weyl
tensor in detail at the linear perturbation level.
 We assume the AdS bulk with the de Sitter brane
as the background. On this background, the 5-dimensional Weyl tensor
vanishes, hence $E_{\mu\nu}=0$.
 We then assume the matter energy momentum tensor on the
brane, $\tau_{\mu\nu}$ to be small and consider $E_{\mu\nu}$ 
induced by $\tau_{\mu\nu}$.

This paper is organized as follows.
In Sec.~II, we derive the general solution to the projected
5-dimensional Weyl tensor perturbation, $E_{\mu\nu}$.
In Sec.~III, we concentrate on the cosmological radion mode and derive
the explicit form of $E_{\mu\nu}$ in terms of the radion after zero-mode
truncation. The physical meaning of the radion is also discussed.
In Sec.~IV, we summarize the result and discuss its implications.

\section{Weyl tensor perturbation}
In this section, we derive the general solution to the projected
5-dimensional Weyl tensor perturbation, $E_{\mu\nu}$.
Although our prime interest is on the
3-brane in the 5-dimensional spacetime, we generalize equations for 
the $(n-1)$-brane in the $(n+1)$-dimensional spacetime unless it is
necessary to restrict to $n=4$.
 The advantage is that it makes easy to see what is
affected by spacetime dimensions and what is not. 

First, we summerize the basic equations of the system with two
($n-1$)-branes as the fixed points of the $Z_2$-symmetry in a
($n+1$)-dimensional spacetime with negative cosmological constant
$\Lambda_{n+1}$. The bulk metric $g_{ab}$ obeys the $(n+1)$-dimensional
Einstein equations:
\begin{eqnarray}
 \label{bulkEin}
  {}^{(n+1)}G_{ab}+\Lambda_{n+1}g_{ab}=\kappa^2\,T_{ab}\,,
\end{eqnarray}
where $\kappa^2$ is the $(n+1)$-dimensional gravitational constant
and $T_{ab}$ is assumed to be zero except on the branes.
We use the latin indices for $(n+1)$-dimensions in the bulk
and the Greek indices for $n$-dimensions on the brane.
On the brane, we decompose the energy momentum tensor into
the tension part, $-\sigma g_{\mu\nu}$ and the matter part,
 $\tau_{\mu\nu}$. Although this decomposition is arbitrary in general,
below we consider the background with $\tau_{\mu\nu}=0$, which 
fixes the decomposition.

As derived in \cite{SMS}, the effective Einstein equations on the brane
is written in the form,
\begin{eqnarray}
 \label{branEin}
{}^{(n)}G_{\mu\nu}+\Lambda_{n} q_{\mu\nu}
=\kappa_{n}^2\tau_{\mu\nu}+\kappa^4\pi_{\mu\nu}-E_{\mu\nu}
\end{eqnarray}
where $q_{\mu\nu}$ is the intrinsic metric on the brane,
\begin{eqnarray}
 \label{branLamb}
  \Lambda_n := \frac{n-2}{n}\kappa^2
     \left( \Lambda_{n+1}+\frac{n}{8(n-1)}\kappa^2\sigma^2\right),\ \ \ 
   \kappa_n := \frac{n-2}{4(n-1)}\sigma \kappa^4,
\end{eqnarray}
$\pi_{\mu\nu}$ is a tensor quadratic in
$\tau_{\mu\nu}$ whose explicit form is unnecessary,
 and 
\begin{eqnarray}
 \label{Edefined}
E_{\mu\nu}:={}^{(n+1)}C^a_{\ \mu b\nu}n_{a}n^{b}
\end{eqnarray}
is the projected $(n+1)$-dimensional Weyl tensor where
$n^a$ is the vector unit normal to the brane.

For appropriate choice of the brane tensions,
with $\tau_{\mu\nu}=\pi_{\mu\nu}=E_{\mu\nu}=0$,
 the Randall-Sundrum (RS) branes\cite{RS1,RS2} and 
the de Sitter branes discussed by Garriga and Sasaki\cite{GS} are 
solutions to these equations. 
The bulk spacetime of these solutions is anti-de Sitter
(AdS$^{n+1}$), and their metric can be concisely expressed as
\begin{eqnarray}
 ds^2&=&a(z)^2\left(dz^2+\gamma_{\mu\nu}dx^\mu dx^\nu \right);
\nonumber
\\  \qquad
&& a(z)={\ell\sqrt{{\cal K}}\over\sinh\sqrt{{\cal K}}z}\,,
\quad \ell=\left[{-n(n-1)\over2\Lambda_{n+1}}\right]^{1/2},
  \label{adsnp1}
\end{eqnarray}
where $\gamma_{\mu\nu}$ is the metric of a Lorentzian
$n$-dimensional constant curvature space with curvature
${\cal K}=0$ or ${\cal K}=+1$, and
$\ell$ is the curvature radius of the AdS$^{n+1}$. 
Then, placing two branes with positive and negative tensions
 at $z=z_{\pm}$  ($0<z_+<z_-$), and gluing two copies of the region 
$z_+ \le z \le z_-$ together, 
we obtain the RS flat branes and the de-Sitter
branes for ${\cal K}=0$ and ${\cal K}=+1$, respectively,
in the AdS bulk.
The tensions on these branes are given by
\begin{eqnarray}
  \sigma_{\pm}
=\pm\frac{2(n-1)}{\kappa^2\ell}\cosh{\sqrt{{\cal K}}z_{\pm}}\,.
\end{eqnarray}
The $n$-dimensional cosmological constant $\Lambda_n$ on each
of the branes is given by
\begin{eqnarray}
\Lambda_n=\frac12{(n-1)(n-2){\cal K}\over a_{\pm}^2}\,,
\end{eqnarray}
where $a_{\pm}=a(z_{\pm})$. 

Now we consider the perturbation of these solutions.
We assume $\tau_{\mu\nu}$ to be of $O(\epsilon)$
and consider the linear response of $E_{\mu\nu}$ to $\tau_{\mu\nu}$.
We can consistently neglect the tensor $\pi_{\mu\nu}$ at the linear
perturbation level.

The linear perturbation equation for $E_{\mu\nu}$
is derived in \cite{SSM} on the RS flat brane background.
Here we generalize it to the de Sitter brane background.
For this purpose, it is useful to introduce the scaled
$E_{\mu\nu}$ by
\begin{eqnarray}
  \label{hatEdef}
  \hat E_{\mu\nu}=a^{n-2}E_{\mu\nu}.
\end{eqnarray}
Then, linearizing the equations given in \cite{SMS},
the equation of
motion for $\hat E_{\mu\nu}$ in the bulk is found as
\begin{eqnarray}
  \label{hatEeq}
  \left[a^{n-3}{\partial\over\partial z}
  {1\over a^{n-3}}{\partial\over\partial z}
+\Box_n-nK\right]\hat E_{\mu\nu}
  =0\,
\end{eqnarray}
where  $\Box_n:=\gamma^{\mu\nu}D_{\mu}D_{\nu}$ is
the $n$-dimensional d'Alembertian 
 with respect to the metric $\gamma_{\mu\nu}$, and $D_{\mu}$ 
is the covariant derivative.
{}From the Israel junction condition and the $Z_2$-symmetry, 
the boundary conditions at the branes are given by
\begin{eqnarray}
 \left.\partial_z\hat E_{\mu\nu}\right|_{z=z_{\pm}}
 &=&\pm {\kappa^2\over2}a_{\pm}^{n-3}
\left(\Sigma^T_{\mu\nu}+\Sigma^S_{\mu\nu}\right)\,,
  \label{bcond}
\end{eqnarray}
where
\begin{eqnarray}
&&  \Sigma^T_{\mu\nu}=\left[-\Box_n+nK\right]
\left(\tau_{\mu\nu}-{a^2\over n}\gamma_{\mu\nu}\tau\right),
\label{stdef}\\
&&\Sigma^S_{\mu\nu}=
-{a^2\over n-1}\left[D_\mu D_\nu
-{1\over n}\gamma_{\mu\nu}\Box_n\right]\tau\,,
\label{ssdef}
\end{eqnarray}
with $\tau:=q^{\mu\nu}\tau_{\mu\nu}=a^{-2}\gamma^{\mu\nu}\tau_{\mu\nu}$.

To solve these equations for $\hat{E}_{\mu\nu}$,
we introduce the Green function which satisfies
\begin{eqnarray}
  \label{tensorG}
&&  \left[-a^{n-3}{\partial\over\partial z}
 {1\over a^{n-3}}{\partial\over\partial z}-\Box_n+n{\cal K}\right]
G_{\mu\nu}(x,z;x',z')^{\alpha\beta}
\nonumber\\
&&\hspace{10em}
=a^{n-3}{\delta^n(x-x')\over\sqrt{-\gamma}}\delta(z-z')
\left(\delta_\mu^{(\alpha}\delta_\nu^{\beta)}
-{1\over n}\gamma_{\mu\nu}\gamma^{\alpha\beta}\right)\,;
\nonumber\\
&&\qquad
 \left.\partial_z G_{\mu\nu}{}^{\alpha\beta}\right|_{z=z_{\pm}}=0.
\end{eqnarray}
The conventional causal boundary condition would be to choose the
retarded Green function. For our discussions below, however, it is
unnecessary to specify the causal boundary condition. Hence
we leave it unspecified.
The only restriction is that all Green functions appearing throughout
the paper should have the same causal boundary condition.
Then, the formal solution is given by
\begin{eqnarray}
  \hat E_{\mu\nu}(x,z)
 &=&\int{d^nx'\sqrt{-\gamma(x')}}
\left[{1\over a(z')^{n-3}}G_{\mu\nu}(x,z;x'z')^{\alpha\beta}
{\partial\over\partial z'}\hat E_{\alpha\beta}(x',z')
\right]^{z_{-}}_{z_{+}}
\nonumber\\
&=&-{\kappa^2\over2}\int{d^nx'\sqrt{-\gamma(x')}}
\Bigl\{G_{\mu\nu}(x,z;x'z')^{\alpha\beta}
\left(\Sigma^T_{\alpha\beta}(x',z')+\Sigma^S_{\alpha\beta}(x',z')
\right)
\Bigr\}^{z_-}_{z_+}\,,
  \label{hatEG}
\end{eqnarray}
where
 $\displaystyle \bigl[Q(z')\bigr]^{z_{-}}_{z_{+}}=Q(z_{-})-Q(z_{+})$
and
 $\displaystyle \bigl\{Q(z')\bigr\}^{z_{-}}_{z_{+}}=Q(z_{-})+Q(z_{+})$.

Because the scale factor $a$ of the bulk metric depends only on $z$ 
and because the branes are located at the fixed values of $z$,
the Green function can be expressed in the factorized form
in terms of appropriately normalized
mode functions. Let $\hat E_{\mu\nu}=\psi_m(z) Y^{(m)}_{\mu\nu}(x)$.
Then Eq.~(\ref{hatEeq}) separates as
\begin{eqnarray}
  \label{modefcns}
&&  \left[-a^{n-3}{d\over dz}
  {1\over a^{n-3}}{d\over dz}+(n-2){\cal K}\right]\psi_m=m^2\psi_m\,;
\left.\quad{d\over dz}\psi_m\right|_{z=z_{\pm}}=0.
\label{psim}\\
&&
\left[-\Box_n+m^2+2{\cal K}\right]Y^{(m)}_{\mu\nu}=0\,;
\quad Y^{(m)}_\mu{}^\mu=D^\nu Y^{(m)}_{\mu\nu}=0.
\end{eqnarray}
The $n$-dimensional mode function $Y^{(m)}_{\mu\nu}(x)$
may be further decomposed in terms of the $(n-1)$-dimensional
spatial eigenfunctions (i.e., spatial harmonic functions).
 Since our interest is
in the $z$-dependence of the mode functions, we 
simply denote the $n$-dimensional mode function
by $Y^{(m,k)}_{\mu\nu}$ where $k$ represents the $(n-1)$ sets of
spatial eigenvalues. Nevertheless, it is worth noting that
the spatial $SO(n)$-symmetry of the $(n-1)$-dimensional space
can be used to separate the components of $Y^{(m,k)}_{\mu\nu}$
by their properties under the symmetry transformations.
One then obtain the scalar, vector and tensor type mode functions 
with respect to the $SO(n)$-symmetry. This corresponds to
the decomposition of the perturbation into the scalar, vector
and tensor perturbations in the context of cosmology on the brane,
i.e., in the sense of the $(n-1)$-dimensions.

{}From Eq.~(\ref{psim}), we see that there is a mode $\psi_m=$constant
with $m^2=(n-2){\cal K}$ that
trivially satisfies the equation. 
Setting $\psi_m=a^{(n-3)/2}f_m$ and
 substituting $a(z)=\ell\sqrt{{\cal K}}/\sinh\sqrt{{\cal K}}z$, we find
\begin{eqnarray}
  \label{fmeq}
  \left[-{d^2\over dz^2}
  +{(n-5)(n-3){\cal K}\over2\sinh^2\sqrt{{\cal K}}z}
+\left({n-1\over2}\right)^2{\cal K}
  \right]f_m=m^2f_m\,.
\end{eqnarray}
Thus when the negative-tension brane is absent (or $z_{-}\to\infty$),
the spectrum will be continuous for $m^2\geq[(n-1)/2]^2{\cal K}$.
Since the mode $m^2=(n-2){\cal K}$ translates to $f_m\propto a^{-(n-3)/2}$,
and it does not have a node, it is the lowest eigenmode.
Since 
\begin{eqnarray}
  \left({n-1\over2}\right)^2-(n-2)=\left({n-3\over2}\right)^2\geq0,
\end{eqnarray}
the mode $m^2=(n-2){\cal K}$ is isolated from the other modes for $n>3$.
Although not necessary for the following arguments, 
we assume there is no mode in the range
$(n-2){\cal K}<m^2<[(n-1)/2]^2{\cal K}$. 
This is the case when $n=4$.

The orthonormality of the eigenfunction $\psi_m(z)$
of Eq.~(\ref{modefcns})
is determined by requiring the equation to be self-adjoint.
This gives
\begin{eqnarray}
&&
\int_{z_+}^{z_-}\!dz\, a^{3-n}\psi_m(z)\psi_{m'}(z)
={N_{m^2}\over\ell^{n-3}}\,\delta_{mm'}
\quad\mbox{for}\quad m^2\geq\left({n-1\over2}\right)^2{\cal K}
\,, \label{psinorm}
\\
&&\int_{z_+}^{z_-}\! dz\, a^{3-n}
={N_{m^2}\over\ell^{n-3}}
\qquad\qquad\mbox{for}\quad m^2=(n-2){\cal K}\,,
\label{psi0norm}
\end{eqnarray}
where $N_{m^2}$ is a normalization constant.
In the case of $n=4$,
\begin{eqnarray}
  \label{4dnorm}
  N_{(n-2){\cal K}}={\cosh\sqrt{{\cal K}}z_{-}
-\cosh\sqrt{{\cal K}}z_{+}\over {\cal K}}\,.
\end{eqnarray}
Then we have
\begin{eqnarray}
  \label{Greenform}
  G_{\mu\nu}(x,z;x',z')^{\alpha\beta}
=\ell^{n-3}\left(
{1\over N_{(n-2){\cal K}}}G^{((n-2){\cal K})}_{\mu\nu}(x,x')^{\alpha\beta}
+\sum_{m\geq m_c}
{\psi_m(z)\psi_m(z')\over N_{m^2}}
G^{(m^2)}_{\mu\nu}(x,x')^{\alpha\beta}\right)\,,
\end{eqnarray}
where $m_c^2=\bigl((n-1)/2\bigr)^2{\cal K}$, and 
$G^{(m^2)}_{\mu\nu}(x,x')^{\alpha\beta}$
is the $n$-dimensional part of the Green function 
satisfying 
\begin{eqnarray}
  \label{ndGreen}
  [-\Box_n+m^2+2{\cal K}]G^{(m^2)}_{\mu\nu}(x,x')^{\alpha\beta}
 ={\delta^n(x-x')\over\sqrt{-\gamma}}
\left(\delta_\mu^{(\alpha}\delta_\nu^{\beta)}
-{1\over n}\gamma_{\mu\nu}\gamma^{\alpha\beta}\right)\,.
\end{eqnarray}
The $n$-dimensional Green function is constructed from
properly normalized $Y^{(m,k)}_{\mu\nu}(x)$, but
we do not give the explicit expression of
$G^{(m^2)}_{\mu\nu}(x,x')^{\alpha\beta}$ 
in terms of $Y^{(m,k)}_{\mu\nu}(x)$, since it is unnecessary for
our purpose. 

Before concluding this section, we note an important implication of
the above result to cosmological perturbations of the de Sitter
brane-universe. 
Since $E_{\mu\nu}$ describes all the extra-dimensional
effect on the brane-world, and since it contains all the scalar, vector
and tensor type perturbations in the sense of cosmological perturbations
as mentioned earlier,\footnote{Note that the terms, scalar, vector
and tensor, here are {\it not\/} used in the $n$-dimensional sense,
but in the $(n-1)$-dimensional sense.}
 we see immediately that the spectrum of the Kaluza-Klein modes
is the same for all the types of cosmological perturbations,
except for the lowest (zero) mode $m^2=(n-2){\cal K}$ on which
we discuss in the next section. This is in agreement with the previous
results on cosmological tensor\cite{GS,LMW} and vector\cite{bmw}
 Kaluza-Klein spectra on the de Sitter brane.

\section{Cosmological radion}\label{CosRad}
In this section,
we consider the so-called zero-mode truncation. 
That is, we neglect 
all the Kaluza-Klein modes $m^2\ge m_c^2=[(n-1)/2]^2{\cal K}$.
In our formalism, this means to consider only the mode 
$m^2=(n-2){\cal K}$. The peculiarity of this mode is that
the corresponding $n$-dimensional
 tranceverse-traceless tensor mode degenerates
with the scalar mode of the eigenvalue $m_s^2=-n{\cal K}$
\cite{KS}.
Thus one expects this mode to play the role of the radion.
In passing, it may be worthwhile to mention that the
tranceverse-traceless tensor mode of the eigenvalue $m^2=0$
degenerates with the transverse vector mode of the eigenvalue
$m_v^2=(n-1)K$. However, since the mode $m^2=0$ is absent
in the spectrum of $E_{\mu\nu}$ for $n>3$, we readily see that
there exists no extra radion-like vector mode in the de Sitter
brane-world.

We approximate the Green function as
\begin{eqnarray}
  \label{zerotrunc}
 G_{\mu\nu}(x,z;x',z')^{\alpha\beta}
={\ell^{n-3}\over N_{(n-2){\cal K}}}
G^{((n-2){\cal K})}_{\mu\nu}(x,x')^{\alpha\beta}\,.
\end{eqnarray}
Then from Eq.~(\ref{hatEG}), $\hat E_{\mu\nu}$ is given by
\begin{eqnarray}
  \label{hatEzero}
\hat E_{\mu\nu}(x)=-{\kappa^2\ell^{n-3}\over2N_{(n-2){\cal K}}}
\int d^nx'\sqrt{-\gamma}
G^{((n-2){\cal K})}_{\mu\nu}(x,x')^{\alpha\beta}
\left\{\Sigma^T_{\alpha\beta}(x')+\Sigma^S_{\alpha\beta}(x')
\right\}^{z_-}_{z_+}\,.
\end{eqnarray}
Incidentally, the truncated $\hat E_{\mu\nu}$ turns out to be
independent of the extra-dimensional coordinate $z$.
Noting the form of the source in Eqs.~(\ref{stdef}) and
(\ref{ssdef}), and the fact that the Green function obeys
\begin{eqnarray}
  [-\Box_n+n{\cal K}]G^{((n-2){\cal K})}_{\mu\nu}(x,x')^{\alpha\beta}
 ={\delta^n(x-x')\over\sqrt{-\gamma}}
\left(\delta_\mu^{(\alpha}\delta_\nu^{\beta)}
-{1\over n}\gamma_{\mu\nu}\gamma^{\alpha\beta}\right)\,,
\end{eqnarray}
we find it is convenient to introduce the following
two auxiliary $n$-dimensional fields.
Let $\phi_{\mu\nu}(x)$ and $\phi(x)$ be the $n$-dimensional
tensor and scalar fields which satisfy
\begin{eqnarray}
 &&\left[-\Box_n+n{\cal K}\right]\phi_{\mu\nu}
=\frac{\kappa^2}{2}\Sigma^S_{\mu\nu}\,,
\label{phimneq}
\\
&&\left[-\Box_n-n{\cal K}\right]\phi=-\frac{\kappa^2 a^2}{2(n-1)}\,\tau\,,
 \label{phieq}
\end{eqnarray}
respectively. Then Eq.~(\ref{phimneq}) is rewritten as
\begin{eqnarray}
  \left[-\Box_n+n{\cal K}\right]\phi_{\mu\nu}
   &=& -\left[D_\mu D_\nu
-{1\over n}\gamma_{\mu\nu}\Box_n\right]\frac{\kappa^2a^2}{2(n-1)}\tau
    = \left[D_\mu D_\nu-{1\over n}\gamma_{\mu\nu}\Box_n\right]
      \left[-\Box_n-n{\cal K}\right]\phi \nonumber\\
   &=& \left[-\Box_n+n{\cal K}\right]
       \left[D_\mu D_\nu-{1\over n}\gamma_{\mu\nu}\Box_n\right]\phi\,.
\end{eqnarray}
Hence, with an appropriate causal
boundary condition imposed on the fields, we obtain  
\begin{eqnarray}
  \label{phirel}
  \phi_{\mu\nu}=
 \left[D_\mu D_\nu-{1\over n}\gamma_{\mu\nu}\Box_n\right]\phi\,.
\end{eqnarray}

Using these $\phi_{\mu\nu}$ and $\phi$,
we obtain from Eq.~(\ref{hatEzero})
\begin{eqnarray}
  \label{hatEfinal}
  \hat E_{\mu\nu}(x)
    &=&-{\kappa^2\ell^{n-3}\over2N_{(n-2){\cal K}}}
        \int d^n\!x'\sqrt{-\gamma}
G^{((n-2){\cal K})}_{\mu\nu}(x,x')^{\alpha\beta}  
\left[-\Box_n+n{\cal K}\right]
           \left\{\tau_{\alpha\beta}
-{1\over n}\gamma_{\alpha\beta}\tau+\frac{2}{\kappa^2}\phi_{\mu\nu}
           \right\}^{z_-}_{z_+}\nonumber\\
    &=&-{\ell^{n-3}\over2N_{(n-2){\cal K}}}
         \left\{\kappa^2\bar{\tau}_{\mu\nu}
          +2\left[D_\mu D_\nu-{1\over n}\gamma_{\mu\nu}\Box_n\right]\phi
         \right\}^{z_-}_{z_+}
\end{eqnarray}
where $\bar{\tau}_{\mu\nu}:=\tau_{\mu\nu}-{a^2\over n}\gamma_{\mu\nu}\tau$ 
and the scalar field $\phi$ satisfies Eq.~(\ref{phieq}).
Thus $\hat{E}_{\mu\nu}$ in the zero-mode truncation is given by the
two parts; one given in terms of $\tau_{\mu\nu}$ directly,
and the other determined by the scalar field $\phi$.
This is our main result.  
For $n=4$ and ${\cal K}=0$, we recover the result obtained by
Garriga and Tanaka\cite{GS}. We may therefore identify $\phi$
(to be precise $\bigl\{\phi(x)\bigr\}^{-}_{+}$)
as the radion that describes the relative displacement of the
branes.

To confirm this identification, we derive the second order
action for the metric perturbation $h_{\mu\nu}$ 
in a gauge $h_{ab}n^b=0$ explicitly in Appendix A.
There the action is expressed in terms of the metric perturbation
$h_{\mu\nu}$ in the bulk and the coordinate displacement of the
brane $\varphi$.
Then specializing the gauge to the Randall-Sundrum gauge
in which $h_{ab}n^b=D^{\mu}h_{\mu\nu}=h^{\mu}{}_{\mu}=0$,
the equation of motion of $\varphi$ is found as
\begin{eqnarray}
  \left[-\Box_n-n{\cal K}\right]\varphi(x)
=-\frac{\kappa^2 a^2}{2(n-1)}\tau\,.
\end{eqnarray}
See Eq.~(\ref{EOMvarphi}).
Thus the scalar field $\phi(x)$ introduced here obeys
exactly the same equation as $\varphi(x)$, 
and can be identified as the displacement 
of the brane in the Randall-Sundrum gauge.

It is instructive to consider the physical meaning of $\phi$
from the view point of an observer on the brane. 
For this purpose, we take the gaussian normal
coordinates around the brane, defined by $h_{ab}n^b=0$ and
$\varphi(x)=0$. The metric perturbation in this gauge
describes what one can observe on the brane.
In this gauge, we may decompose the metric perturbation as
\begin{eqnarray}
  \label{hmngauss}
 h_{\mu\nu}=a^2[{\cal R}\gamma_{\mu\nu}+D_{\mu}D_{\nu}H_T
+D_{(\mu}X_{\nu)}+X_{\mu\nu}]\,,
\end{eqnarray}
where $X_{\mu}$ is transverse; $D^{\mu}X_{\mu}=0$,
 and $X_{\mu\nu}$ is transverse-traceless;
 $D^{\mu}X_{\mu\nu}=X^{\mu}{}_{\mu}=0$.
The $n$-dimensional scalar potential ${\cal R}$ describes the
scalar curvature perturbation of the brane\cite{KS}.
Then the equation of motion of $\varphi(x)$, Eq.~(\ref{EOMvarphi}),
reduces to
\begin{eqnarray}
  \left[-\Box_n-n{\cal K}\right]{{\cal R}\over H}
=-\frac{\kappa^2a^2}{2(n-1)}\tau\,,
   \label{EOMvpGN}
\end{eqnarray}
where $H=-\ppp_z a(z)/a^2(z)$.
Thus $\phi(x)$ in Sec.~\ref{CosRad} obeys exactly
 the same equation as ${\cal R}/H$ in the gaussian normal coordinates.
That is, $\phi$ itself is not an additional (gravitational)
scalar field on the brane, but simply corresponds to the
intrinsic scalar curvature of the brane.
It is important to note here that the above equation (\ref{EOMvpGN}) is
nothing but the trace of the linear perturbation of the effective
Einstein equations (\ref{branEin}) in the present background spacetime.

Thus it is not quite adequate to call $\phi$ the radion.
The term, radion, should be used to describe the combined effect
of the intrinsic scalar curvature perturbation of
the brane and its effect on the non-trivial Weyl curvature perturbation
$E_{\mu\nu}$ in the bulk which reacts back on the brane.
In the two-brane system, this is indeed the case, and
the scalar field $\phi$ concisely describes the effect.

For the single positive-tension brane case, we just need to discard the
source term on the negative-tension brane in the result
Eq.~(\ref{hatEfinal}) and take the limit
$z_{-}\rightarrow\infty$ in the above argument. 
Note that $z\to\infty$ corresponds to the center of the $(n-1)$-sphere
in the bulk $n$-dimensional spatial hypersurface in the case of
${\cal K}=1$. See Eqs.~(\ref{psinorm}) and (\ref{psi0norm}). 
Then, for $n=4$ case, we find that 
$N_{(n-2){\cal K}}\rightarrow\infty$ from Eq.~(\ref{4dnorm}), and thus
$\hat E_{\mu\nu}\rightarrow 0$ from Eq.~(\ref{hatEfinal}). 
That is, if there is only one positive-tension brane,
the radion is absent and the gravity on the de Sitter brane reduces to
the conventional 4-dimensional gravity with corrections solely from the
Kaluza-Klein modes.

\section{Summary and Discussion}

In this paper, we have investigated the role of the radion mode on the
de Sitter brane, using the covariant curvature tensor formalism
developed in \cite{SMS}. We have found that if there is only one
positive-tension de Sitter brane, the Einstein gravity is recovered on
the brane without any corrections other than those from the Kaluza-Klein
modes. There is no trace of the radion mode, at least at the linear
order, contrary to our 
naive expectation that the displacement of the brane does have physical
meaning for the de Sitter 3-sphere brane and hence that some trace of it
should appear on the brane. If we recall the case of
a vacuum bubble in the 4-dimensional spacetime\cite{GarVil}, we find
exactly the same phenomenon there; an observer on the wall cannot
detect fluctuations of the wall, though the brane does fluctuate
in the 4-dimensional sense. 

To investigate the physics behind this phenomenon further,
as well as to clarify the role of the scalar field $\phi$,
we have derived the action for the displacement of the brane, $\varphi$,
coupled to the metric perturbation, $h_{\mu\nu}$,
in Appendix A. The equation of motion of
$\varphi$ is then found to be exactly the same as that of
$\phi$ in the Randall-Sundrum gauge.
This result may be interpreted as follows. The brane 
does fluctuate in this gauge by the amount $\varphi(x)$
in the $(n+1)$-dimensional sense.
However, the displacement of the brane cannot be detected
by itself because one can always choose a local coordinate
system in which $\varphi=0$. 
{}From the view point of an observer on the brane, 
the natural coordinate associated with it is the gaussian normal
coordinates defined by $h_{ab}n^b=0$ and $\varphi=0$. 
We have then found $\phi$ describes the intrinsic $n$-dimensional
scalar curvature perturbation of the brane.
Hence the effect due to the fluctuation of the brane 
cannot be detected unless it causes a non-trivial metric perturbation
in the bulk and propagates back to the brane (or to the other brane
in the two-brane case).
Since the physical metric perturbation in the bulk is described
by $E_{\mu\nu}$, the fact that the Green function in
the zero-mode truncation, Eq.~(\ref{zerotrunc}),
 vanishes in the limit of a single brane 
implies that the bulk metric becomes increasingly rigid as the
location of the negative-tension brane approaches $z_{-}\to\infty$
and the fluctuation ceases to propagate through the bulk. As a result,
the gravity on the positive-tension brane is not affected by the
fluctuation of the brane, $E_{\mu\nu}=0$, in the single brane case. 

In the two-brane system, on the other hand, there exists
the cosmological radion mode
as in the flat two-brane system. An interesting
fact is that the cosmological radion has the negative mass-squared
$m^2=-n{\cal K}$. This may have important implications
to the cosmological perturbation theory.
In connection with this issue, 
we should mention that our result that the radion has the negative
mass-squared is in apparent contradiction with the result
obtained by Chiba\cite{Chiba}, who derived a non-linear effective action
for the radion mode by assuming a certain form of the 5-dimensional
metric. Naively applying his result to curved background and taking
the large separation limit, 
one finds the radion is a massless conformal scalar, i.e., it
effectively has a positive mass-squared on the de Sitter brane. However,
it should be noted that the assumed form of the 5-dimensional action
used in \cite{Chiba} is valid only for the flat brane background. So,
there is in fact no logical contradiction. To understand the role of the
radion in the inflationary brane universe, it will be necessary to
derive the effective action of the radion on the de Sitter brane.
This issue is currently under investigation.

\section*{Acknowledgements}
This work is supported in part by the Yamada Science Foundation
and by Monbusho Grant-in-Aid for Scientific Research, No.~12640269.
We are grateful to Jaume Garriga for stimulating discussions.
UG would like to thank Tetsuya Shiromizu for his help at the early
stage of the work.

\appendix

\section{the surface term action for a thin-wall system}

In this appendix, we consider an $n$-dimensional thin-wall 
(i.e., $(n-1)$-brane) system in the $(n+1)$-dimensional bulk spacetime
and derive the perturbative second order action for the displacement
of the brane.
The system consists of two manifolds ${\cal M}_+$ and ${\cal M}_-$ with a
common boundary $\ppp {\cal M}_+=\ppp {\cal M}_-=\Sigma$ on which
the tension $\sigma$ and matter fields with the Lagrangian ${\cal L}_m$
are present. 
The background metric is assumed to be of the form,
$ds^2=dr^2+q_{\mu\nu}dx^\mu dx^\nu$ with the surface $\Sigma$ at
$r=r_0$.
For most of the discussions below, we shall not impose
the $Z_2$-symmetry across $\Sigma$.
Hence, the result can be applied to a general thin-wall system.

The action of the system is given by
\begin{eqnarray}
 \label{L}
  && I = I_{R\mbox{-}2\Lambda,+}+I_{R\mbox{-}2\Lambda,-}
+I_{K,+}-I_{K,-}+I_\sigma +I_m\,;\nnb\\
  && \hspace{7ex} I_{R\mbox{-}2\Lambda,\pm}:=
       \frac{1}{2\kappa^2}\int_{{\cal M}_{\pm}}
\hspace{-2ex}\sqrt{-g} dr d^n\! x\, (R-2\Lambda)\,,
     \hspace{3ex} I_{K,\pm}:=
       \frac{1}{\kappa^2}\int_{\ppp{\cal M}_+}
\hspace{-2.5ex}\sqrt{-q}d^n\!x\, K\,,\nnb\\
  && \hspace{7ex} I_\sigma:=-\int_{\Sigma}
\hspace{-0.5ex}\sqrt{-q}d^n\!x\, \sigma\,,
     \hspace{4ex} I_m:=\frac12\int_{\Sigma}
\hspace{-0.5ex}\sqrt{-q}d^n\!x\, {\cal L}_m\,,
\nnb
\end{eqnarray}
where $q_{ab}=g_{ab}-n_a n_b$, 
$K_{ab}:=q_a^{\ c}q_b^{\ d}\V_c n_d$ and $n^a$ is defined such that it
points from ${\cal M}_+$ to ${\cal M}_-$.
The variation of this action with respect to the bulk gravitational
field gives the Einstein equations in 
${\cal M}_+$ and ${\cal M}_-$ and that with respect to the
metric on $\Sigma$ gives the Israel junction condition.
Our strategy to obtain the second order action is as follows.
Apart from simple terms which can be expanded to second order
easily, we consider the surface term $\delta I_\Sigma$
arising from the variation of the action by assuming the bulk
gravitational equations are satisfied. Then considering the displacement
 $r_0\to r_0\pm\varphi_{\pm}(x)$ of the boundary surface
and the metric perturbation $q_{\mu\nu}\to q_{\mu\nu}+h_{\mu\nu}$,
 we integrate the variation $\delta I_\Sigma$ to
obtain the second order action on $\Sigma$.

The action may be rewritten as
\begin{eqnarray}
  &&I=\bigg\{\frac{1}{2\kappa^2}\int_{{\cal M}}
\hspace{-2ex} drd^n\!x\, {\cal I}_{R\Lambda}\bigg\}^+_-
     +\frac{1}{\kappa^2}\int_{\Sigma}
\hspace{-1ex} d^n\!x\, {\cal I}_{K\sigma}+I_m\,;\nnb\\
  && \hspace{8ex} 
  {\cal I}_{R\Lambda}:=\sqrt{-g}\Big[R-2\Lambda\Big]\,,\ \ \ 
  {\cal I}_{K\sigma}:=\sqrt{-q}\Big[\Delta K-\sigma\kappa^2\Big]\,,
\end{eqnarray}
where $\big\{\int_{\cal M}~\big\}^+_-=\int_{\cal M_+}~+\int_{\cal M_-}~$,
$\Delta Q:=Q_+-Q_-$ and $Q_\pm$ is a quantity on $\Sigma$
calculated from the ${\cal M}_\pm$ sides, respectively.
The variation of the action is given by
\begin{eqnarray}
  &&\delta I
      =\bigg\{\frac{1}{2\kappa^2}\int_{{\cal M}}\hspace{-2ex}dr\,d^n\!x\,
   \frac{\delta{\cal I}_{R\Lambda}}{\delta g_{ab}}\,
 \delta g_{ab}\bigg\}^+_-
     +\bigg\{\frac{1}{2\kappa^2}
         \int_{\Sigma}\hspace{-1ex}d^n\!x\,
                   {\cal I}_{R\Lambda}\delta r\bigg\}^+_-\nnb\\
  &&\hspace{10ex}
      +\frac{1}{\kappa^2}\int_{\Sigma}\hspace{-1ex} d^n\!x\,
    \Big[\Delta\pi^{ab}+\frac{\delta{\cal I}_{K\sigma}}
            {\delta q_{ab}}\Big]\delta q_{ab}
      +\frac{1}{\kappa^2}\delta_\Sigma
       \left(\int_{\Sigma}\hspace{-1ex} d^n\!x\, {\cal I}_{K\sigma}\right)
             \nnb\\
  &&\hspace{10ex}
      +\frac12\int_{\Sigma}
\hspace{-1ex}\sqrt{-q} d^n\!x\, \tau^{\mu\nu}\delta q_{\mu\nu}
      +\frac12\delta_\Sigma\left(\int_{\Sigma}
\hspace{-1ex}\sqrt{-q} d^n\!x\, {\cal L}_m\right),
     \label{deltaI}
\end{eqnarray}
where $\delta r$ describes the variation of the
domain of integration $\delta{\cal M}$,
 $\pi^{ab}$ is the boundary term that appears from 
${\cal I}_{R\Lambda}$ by the variation of the metric,
 and $\delta_\Sigma$ denotes the variation 
with respect to the displacement of the boundary surface
without changing the metric on the manifolds ${\cal M}_{\pm}$.
 Note here 
\begin{eqnarray}
   \Delta\pi^{ab}+\frac{\delta{\cal I}_{K\sigma}}{\delta q_{ab}}
     =\Delta K^{ab}-\Delta K q^{ab}+\sigma\kappa^2q^{ab}.
\label{pieq}
\end{eqnarray}

Now we consider the perturbation. We choose the `synchronous'
coordinates with respect to the extra-dimensional direction. So, 
the metric in ${\cal M}_{\pm}$ takes the form,
\begin{equation} 
ds^2_{\pm}=dr^2_{\pm}+(\bar q_{\mu\nu}+h_{\mu\nu})dx^\mu dx^\nu\,,
\end{equation}
in which the unperturbed boundary $\bar \Sigma$ is at $r_{\pm}=r_0$ and
both $r_{\pm}$ increase towards $\bar \Sigma$,
$\bar q_{\mu\nu}$ is the background metric, and $h_{\mu\nu}$
is the perturbation of the metric. Here and in what follows,
barred quantities denote background quantities.
The perturbed boundary $\Sigma$ is placed at
$r_{\pm}=r_0\pm\varphi_{\pm}(x)$.
The perturbation of the surface term is most conveniently 
calculated by introducing the new coordinates
defined by $\hat r:=r\mp\varphi_{\pm}(x)$ and $\hat x^\mu:=x^\mu$ in which
the $\hat r$-coordinate of the surface $\Sigma$ is unperturbed
at $\hat r=r_0$.
Then the metric takes the form,
\begin{eqnarray}
 && ds^2=d\hat r^2+2\varphi_{|\mu}d\hat r d\hat x^\mu
        +\left[\bar q_{\mu\nu}+h_{\mu\nu}
+\varphi_{|\mu}\varphi_{|\nu}\right]d\hat x^\mu d\hat x^\nu\,,
 \nnb\\
 && g^{ab}=(1+\varphi_{|\rho}\varphi^{|\rho})(\ppp_{\hat r})^a(\ppp_{\hat r})^b 
 -2[\varphi^{|\mu}-h^{\mu\nu}\varphi_{|\nu}](\ppp_{\hat r})^{(a}(\ppp_\mu)^{b)}
    +\bar q^{\mu\nu}(\ppp_\mu)^a(\ppp_\nu)^b\,.
\end{eqnarray}
Hence, the vector $n^a$ unit normal to the boundary $\Sigma$ at
$\hat r=r_0$ is given by 
\begin{eqnarray}
  n^a = (1+\frac12\varphi_{|\rho}\varphi_{|\rho})(\ppp_{\hat r})^a
        -[\varphi^{|\mu}-h^{\mu\nu}\varphi_{|\nu}](\ppp_\mu)^{a}.
\label{rhatn}
\end{eqnarray}
The metric on the boundary and its determinant are given by
\begin{eqnarray}
  ds^2_\Sigma&=&[\bar q_{\mu\nu}+h_{\mu\nu}
+\varphi_{|\mu}\varphi_{|\nu}]d\hat x^\mu d\hat x^\nu\,,
\nnb\\
  \sqrt{-q(\hat x)}&=& \sqrt{-\bar q}
        \left[
               1+\frac12(h+\varphi_{|\rho}\varphi^{|\rho})
                +\frac18(h^2-2h^{\rho\sigma}h_{\rho\sigma})
       \right].
\label{rhatq}
\end{eqnarray}
The extrinsic curvature of the boundary surface is given by
\begin{eqnarray}
    K_{\mu\nu}
     &=& \bar K_{\mu\nu}+k_{\mu\nu}-\varphi_{|(\mu\nu)}
     +\frac12\varphi_{|\rho}\varphi^{|\rho}\bar K_{\mu\nu}\,;
   \ \ \ k_{\mu\nu}:=\frac12\ppp_r h_{\mu\nu}\,,
\label{Kexpanded}
\end{eqnarray}

Now, we calculate the perturbation of the surface term
in Eq.~(\ref{deltaI}).
The first term in Eq.~(\ref{deltaI}) does not contribute to the surface
term. The second term gives
\begin{eqnarray}
  I_{\delta {\cal M}}
  :=&&\bigg\{
\frac{1}{2\kappa^2}\int_{\delta r}\hspace{-2ex} dr
\int_{\Sigma}d^n\!x\,
      {\cal I}_{R\Lambda}\bigg\}^+_-
\nnb\\
  =&& \frac{1}{2\kappa^2}\int_{r_0}^{r_0+\varphi_+}
\hspace{-4ex}dr\,d^n\!x \,
       \sqrt{-\bar g}\left[\bar R-2\Lambda\right]
     +\frac{1}{2\kappa^2}\int_{r_0}^{r_0-\varphi_-}
\hspace{-4ex}dr\,d^n\!x\, 
       \sqrt{-\bar g}\left[\bar R-2\Lambda\right]
\nnb\\
  =&& \frac{1}{\kappa^2}\int_{\bar \Sigma}\hspace{-1ex}d^n\!x\sqrt{-\bar q}
               \left[
          \frac{2\Lambda}{n-1}
   \left(\Delta\varphi+\frac12 {\cal K}[\varphi_+^2+\varphi_-^2] \right)
              \right]\,,
 \label{VolInc}
\end{eqnarray}
in which the trace of the background Einstein equations
$\bar{R}=2\Lambda(n+1)/(n-1)$ is used.
The third term is the term that gives the 
Israel junction condition in the absence of matter fields. See
Eq.~(\ref{pieq}). Hence, by considering
the perturbed metric (\ref{rhatq}) and extrinsic curvature
(\ref{Kexpanded}), and identifying 
$\delta q_{\mu\nu}=\delta h_{\mu\nu}$,
we find
\begin{eqnarray}
 \delta_h I^{(2)}_\pi:=&&\int_{\Sigma}\hspace{-1ex} d^n\!x 
     \Big[\Delta\pi^{ab}
       +\frac{\delta{\cal I}_{K\sigma}}{\delta q_{ab}}\Big]\delta h_{ab}
 \nnb\\
    =&& \delta_h \bigg\{ 
         -\frac{1}{2}\!\int_{\Sigma}\hspace{-1ex} d^n\!x\sqrt{-\bar q}
         \bigg(\Big[\Delta \bar K^{ab}-\Delta \bar K \bar q^{ab}
             +\sigma\kappa^2\bar q^{ab}\Big]h_{\mu\nu}
                     +{}^{(2)}{\cal I}_J[h,\varphi]\bigg)\bigg\}\nnb\\
     =&&\delta_h \bigg\{ 
         -\frac{1}{2}\int_{\bar \Sigma}\hspace{-1ex} d^n\!x\sqrt{-\bar q}
          \bigg(h_{\mu\nu}\Delta(C_1^{\mu\nu}\varphi)
            +{}^{(2)}{\cal I}_J[h,\varphi]\bigg)\bigg\}\,,\nnb
\end{eqnarray}
where
\begin{eqnarray}
 {}^{(2)}{\cal I}_J[h,\varphi]:=&&
          \Delta k^{\mu\nu}h_{\mu\nu}-\Delta k_{\sigma}^{\ \sigma}h
         +\frac{\sigma\kappa^2}{2(n-1)}(h^2-h^{\mu\nu}h_{\mu\nu})
         -\Delta\varphi^{|(\mu\nu)}h_{\mu\nu}+h\Box\Delta\varphi,\nnb\\
     C_1^{\mu\nu}:=&&
       \ppp_{r}\bar K^{\mu\nu}
        -\bar q^{\mu\nu}\ppp_{r}\bar K +\frac2n\bar K \bar K^{\mu\nu}\,,
\end{eqnarray}
and $\Box=\bar q^{\mu\nu}D_\mu D_\nu$.
The $C_1$ term comes from the displacement of the surface $\Sigma$
from $\bar \Sigma$ in the original coordinates $(r,x^\mu)$.
Thus we obtain
\begin{eqnarray}
  I^{(2)}_\pi
   =-\frac{1}{2\kappa^2}\int_{\bar \Sigma}\hspace{-1ex} d^n\!x\sqrt{-\bar q}
          \bigg(h_{\mu\nu}\Delta(C_1^{\mu\nu}\varphi)
        +{}^{(2)}{\cal I}_J[h,\varphi]\bigg)\bigg\}.
     \label{JuncVar}
\end{eqnarray}
The fourth term is the term arising from the displacement of the
boundary surface $\Sigma$ without altering the metric on the
Manifold. This results in the changes 
$\bar{q}_{\mu\nu}\to\bar{q}_{\mu\nu}+\varphi_{|\mu}\varphi_{|\nu}$
and $\bar{n}^a\rightarrow\bar{n}^a-\varphi^{|\mu}(\ppp_\mu)^a$,
in addition to the change in the $r$-coordinate of $\Sigma$.
Hence, the fourth term gives
\begin{eqnarray}
 I_{\delta\Sigma}:=&&
   {1\over\kappa^2}
\left(\int_{\Sigma}\hspace{-1ex} d^n\!x {\cal I}_{K\sigma}
      - \int_{\bar \Sigma}\hspace{-1ex} d^n\!x
         \bar {\cal I}_{K\sigma}\right)
\nnb\\
  =&&{1\over\kappa^2}\int_{\Sigma}\hspace{-1ex} d^n\!x\sqrt{-\bar q}
              \Big[-\frac12\sigma\kappa^2(\varphi_+\Box\varphi_+ 
+\varphi_-\Box\varphi_-)\Big]
\nnb\\
 &&\hspace{20ex}
      +{1\over\kappa^2}
       \left(\int_{\Sigma}\hspace{-1ex} d^n\!x\sqrt{-\bar q}
                      \Big[\Delta\bar K -\sigma\kappa^2\Big]
                   -\int_{\bar \Sigma}\hspace{-1ex} d^n\!x\sqrt{-\bar q}
                      \Big[\Delta\bar K -\sigma\kappa^2\Big]
            \right)
\nnb\\
 =&&{1\over\kappa^2}
   \int_{\bar \Sigma}\hspace{-1ex} d^n\!x\sqrt{-\bar q}
         \Big[-\frac12\sigma\kappa^2
            (\varphi_+\Box\varphi_+ +\varphi_-\Box\varphi_-)
   +\Delta(C_2\varphi)+\frac12(C_{3,+}\varphi_+^2 +C_{3,+}\varphi_-^2)
           \Big]\,;\\
 && C_2:=\frac1n \bar K^2 + \ppp_{r}\bar K\,,\quad
      C_3:=\frac1{n}\Big[(2n+1)\bar K\ppp_{r} \bar K+\bar K^3+n\ppp_{r}^2 \bar K\Big],
     \label{KsVar}
\end{eqnarray}
where Eq.~(\ref{Kexpanded}) with $h_{\mu\nu}=0$ has been also used.
As the $C_1$ term, $C_2$ and $C_3$ terms come from
the displacement of the surface $\Sigma$
in the original coordinates $(r,x^\mu)$.
The fifth and the sixth terms give 
\begin{eqnarray}
   I^{(2)}_m:=&&
 \frac12\int_{\bar \Sigma}\hspace{-1ex}\sqrt{-q} d^n\!x
    \tau^{\mu\nu}h_{\mu\nu}
  +\frac12\left(\int_{\Sigma}\hspace{-1ex}\sqrt{-q} d^n\!x 
              {\cal L}_m
         -\int_{\bar \Sigma}\hspace{-1ex}\sqrt{-\bar q} d^n\!x 
              {\cal L}_m\right)
\nnb\\
    =&& \frac12\int_{\bar \Sigma}\hspace{-1ex}
             \sqrt{-\bar q}d^n\!x \tau^{\mu\nu} h_{\mu\nu}
           +\frac12\int_{\bar \Sigma}
            \frac12\Delta
     \left(\frac{\delta(\sqrt{-q}d^n\!x{\cal L}_m)}{\delta \bar q_{\mu\nu}}
                 \frac{\ppp \bar q_{\mu\nu}}{\ppp r} \varphi
             \right)
\nnb\\
    =&& \frac12\int_{\bar \Sigma}\hspace{-1ex}\sqrt{-\bar q}d^n\!x
               \Big[
       \tau^{\mu\nu} h_{\mu\nu}+\tau^{\mu\nu}\Delta(\bar K_{\mu\nu}\varphi)
               \Big]\,.
  \label{I_m}
\end{eqnarray}
Adding up Eqs.~(\ref{VolInc}), (\ref{JuncVar}), (\ref{KsVar}) and
(\ref{I_m}) all together, we find the action
for the surface term, $I_\Sigma[h,\varphi]$, up to second order as
\begin{eqnarray}
 I_\Sigma[h,\varphi]=&&I_{\delta{\cal M}}+I_{\delta\Sigma}
                      +I^{(2)}_\pi+I^{(2)}_m
\nnb\\
 =&&I^{(1)}_\Sigma[h,\varphi]+I^{(2)}_\Sigma[h,\varphi]\,;
\nnb\\
  &&\hspace{3ex}
   I^{(1)}_\Sigma[h,\varphi]
    :=\frac{1}{\kappa^2}\int_{\bar \Sigma}\hspace{-1ex} d^n x \sqrt{-\bar q}
 \left[\Delta(C_2\varphi) + \frac{2\Lambda}{n-1}\Delta\varphi\right]\,,\\
  &&\hspace{3ex}
   I^{(2)}_\Sigma[h,\varphi]
    :=\frac{1}{2\kappa^2}\int_{\bar \Sigma}\hspace{-1ex} d^n x \sqrt{-\bar q}
       \bigg[ - \Delta k^{\mu\nu}h_{\mu\nu}+ \Delta k_{\sigma}^{\ \sigma}h 
      - \frac{\sigma\kappa^2}{2(n-1)}(h^2-h^{\mu\nu}h_{\mu\nu})
\nnb\\
  &&\hspace{25ex}
     +h_{\mu\nu}\Delta \varphi^{|\mu\nu}
     - h\Box\Delta \varphi-h_{\mu\nu}\Delta(C_1^{\mu\nu}\varphi)
             +\kappa^2\tau^{\mu\nu}\Delta(\bar K_{\mu\nu}\varphi)
      + \kappa^2 \tau^{\mu\nu} h_{\mu\nu}\nnb\\
  &&\hspace{25ex}
   -\frac12\sigma\kappa^2
       (\varphi_+\Box\varphi_+ +\varphi_-\Box\varphi_-)
          +(C_{3,+}\varphi_+^2+C_{3,-}\varphi_-^2)
           +\frac{2\Lambda}{n-1}{\cal K}(\varphi_+^2+\varphi_-^2)
           \bigg]\,.
\end{eqnarray}

For the brane-world system specified by Eq.~(\ref{adsnp1}) with 
$Z_2$-symmetry, we have
\begin{eqnarray}
  &&\bar K^{\mu\nu}:=\bar K^{\mu\nu}_+=-\bar K^{\mu\nu}_-
    =\frac12 \Delta \bar K_{\mu\nu}\,,\quad
 \varphi:=\varphi_+=-\varphi_-=\frac12 \Delta \varphi \nnb\\
  &&\bar K^{\mu\nu}= H q^{\mu\nu}\,,\quad
  \ppp_r H =-a^{-2}{\cal K}\,,\quad
  \ppp_r^2 H=2Ha^{-2}{\cal K}\,,\quad
  2\Lambda=-\ell^{-2}n(n-1)\,,\nnb
\end{eqnarray}
where $H=\partial_r a/a$, and hence,
\begin{eqnarray}
&&C_1^{\mu\nu}=2(n-1)a^{-2}{\cal K}q^{\mu\nu}\,,
\quad C_2=-4\Lambda/(n-1)\,,
\nnb\\
&&C_{3,\pm}=2n^2H\ell^{-2}-n\sigma\kappa^2 a^{-2}{\cal K}\,.
\end{eqnarray}
As a result, we have $I^{(1)}_\Sigma=0$,
and $I^{(2)}_\Sigma$ reduces to
\begin{eqnarray}
 I^{(2)}_\Sigma[h,\varphi]
      &=&\frac{1}{\kappa^2}\int_{\Sigma}\hspace{-1ex} d^n x \sqrt{-q}
          \bigg[ - k^{\mu\nu}h_{\mu\nu}+ k_{\sigma}^{\ \sigma}h 
                 - \frac12H(h^2-h^{\mu\nu}h_{\mu\nu})\nnb\\
      &&\hspace{17ex}
             +h_{\mu\nu}\varphi^{|\mu\nu}
- ha^{-2}\Box_n\varphi-(n-1)a^{-2}{\cal K}h\varphi 
 + \frac12\kappa^2 \tau^{\mu\nu} h_{\mu\nu}\nnb\\
      &&\hspace{17ex}
             +\frac12a^{-2}\sigma\kappa^2
   \left(-\varphi\Box_n\varphi-n{\cal K}\varphi^2
    +\frac{\kappa^2a^2}{n-1}\tau\varphi\right)
           \bigg]\,,
\end{eqnarray}
where $\Box_n=a^2\Box=\gamma^{\mu\nu}D_\mu D_\nu$.

Taking the variation of the action $I^{(2)}_\Sigma[h,\varphi]$
with respect to $\varphi$, we obtain
\begin{eqnarray}
 \label{EOMvarphi}
  \left[-\Box_n-n{\cal K}\right]\varphi
=-\frac{\kappa^2a^2}{2(n-1)}\tau+{\cal Q}[h]\,,
\end{eqnarray}
where
\begin{eqnarray}
     {\cal Q}[h]:=\frac1{\sigma\kappa^2}
\left[-a^2 h_{\mu\nu}^{\ \ |\mu\nu}+(\Box_n +(n-1){\cal K})h\right]\,.
\end{eqnarray}
This equation of motion shows how the displacement
$\varphi$ is coupled with the metric and matter fields on the brane.
Taking the variation of the action $I^{(2)}_\Sigma[h,\varphi]$
with respect to $h_{\mu\nu}$, we obtain
\begin{eqnarray}
 - k^{\mu\nu}+H h^{\mu\nu}
     +k_{\sigma}^{\ \sigma}q^{\mu\nu}-Hhq^{\mu\nu}
     + \varphi^{|\mu\nu}+a^{-2}{\cal K}\varphi q^{\mu\nu}
     - a^{-2}q^{\mu\nu}(a^{2}\Box_n+n{\cal K})\varphi
     +\frac12\kappa^2\tau^{\mu\nu}=0\,.
\end{eqnarray}
By subtracting off the trace of the above equation and using
Eq.~(\ref{EOMvarphi}), we find
\begin{eqnarray}
\ppp_r h_{\mu\nu}-2Hh_{\mu\nu}
  = 2\left[\varphi_{|\mu\nu}+{\cal K}\varphi \gamma_{\mu\nu}\right]
     + \kappa^2\left(\tau_{\mu\nu}-\frac1{n-1}\tau q_{\mu\nu} \right)\,.
\end{eqnarray}
This is the perturbed junction condition in the `synchronous' gauge;
$h_{ab}n^b=0$.

\newcommand{\np}{Nucl. Phys. }


\begin{thebibliography}{22}
\bibitem{SMS}   T. Shiromizu, K. Maeda and M. Sasaki,
  \prd {\bf 62}, 024012 (2000).

\bibitem{Antoniadis}
I.~Antoniadis,
Phys.\ Lett.\  {\bf B246}, 377 (1990);\\
J.~Polchinski,
Phys.\ Rev.\ Lett.\  {\bf 75}, 4724 (1995) [hep-th/9510017];\\
%
P.~Horava and E.~Witten,
Nucl.\ Phys.\  {\bf B460}, 506 (1996) [hep-th/9510209];\\
%
N.~Arkani-Hamed, S.~Dimopoulos and G.~Dvali,
Phys.\ Lett.\  {\bf B429}, 263 (1998) [hep-ph/9803315];\\
%
A.~Lukas, B.~A.~Ovrut and D.~Waldram,
Phys.\ Rev.\  {\bf D60}, 086001 (1999)
[hep-th/9806022].

\bibitem{RS1}   L. Randall and R. Sundrum, \prl {\bf 83}, 4690 (1999).
\bibitem{RS2}   L. Randall and R. Sundrum, \prl {\bf 83}, 3370 (1999).
\bibitem{GT}    J. Garriga and T. Tanaka, \prl {\bf 84}, 2778 (2000).
\bibitem{CGR}   C.~Charmousis, R.~Gregory and V.A.~Rubakov,
 \prd  {\bf 62}, 067505 (2000).
\bibitem{GS}    J. Garriga and M. Sasaki, \prd {\bf 62}, 043523 (2000).

\bibitem{Mukohyama}
S.~Mukohyama,
Phys.\ Rev.\  {\bf D62}, 084015 (2000)
[hep-th/0004067];
ibid., hep-th/0006146;
%
\bibitem{Kodama}
H.~Kodama, A.~Ishibashi and O.~Seto,
Phys.\ Rev.\ D {\bf 62}, 064022 (2000) [hep-th/0004160];
%
\bibitem{HHR}
S.W. Hawking, T. Hertog, and H.S. Reall, Phys. Rev. D {\bf 62},
043501 (2000) [hep-th/0003052];  hep-th/0010232.
%
\bibitem{Carsten}
C.~van de Bruck, M.~Dorca, R.~H.~Brandenberger, and A.~Lukas,
Phys. Rev. D {\bf 62}, 123515 (2000) [hep-th/0005032];\\
C.~van de Bruck, M.~Dorca, C.~J.~Martins, and M.~Parry,
Phys. Lett. B, to appear [hep-th/0009056].
%
\bibitem{Koyama}
K.~Koyama and J.~Soda, Phys. Rev. D {\bf 62}, 123502 (2000)
[hep-th/0005239].

\bibitem{Langlois}
D.~Langlois, Phys. Rev. D {\bf 62}, 126012 (2000)
[hep-th/0005025];
hep-th/0010063.

\bibitem{MWBH}
R.~Maartens, D.~Wands, B.~A.~Bassett, and I.~Heard,
Phys.\ Rev. D {\bf 62}, 041301 (2000)
[hep-ph/9912464].

\bibitem{LMW}
D.~Langlois, R.~Maartens, and D.~Wands,
Phys.\ Lett.\  {\bf B489}, 259 (2000)
[hep-th/0006007].

\bibitem{bmw}
H. Bridgman, K. Malik, and D. Wands, hep-th/0010133.

\bibitem{cll}
E. Copeland, J. Lidsey, and A. Liddle, astro-ph/0006421.

\bibitem{DDK}
N. Deruelle, T. Dolezel, and J. Katz, hep-th/0010215.

\bibitem{NeroSachs}
A. Neronov and I. Sachs, hep-th/0011254.

\bibitem{Roy}
R.~Maartens,
Phys.\ Rev.\ D {\bf 62}, 084023 (2000)
[hep-th/0004166];\\
C.~Gordon and R.~Maartens, Phys. Rev. D, to appear
[hep-th/0009010].

\bibitem{SSM}   M. Sasaki, T. Shiromizu and K. Maeda, \prd {\bf 62}, 
024008 (2000).
\bibitem{KS} H. Kodama and M. Sasaki, Prog. Theor. Phys. Suppl.
{\bf 78}, 1 (1984).
 \bibitem{GarVil} J.~Garriga and A.~Vilenkin, \prd {\bf 44}, 1007
 (1991); \prd {\bf 45}, 3469 (1992).
 \bibitem{Chiba}     T. Chiba, \prd {\bf 62}, 021502, (2000).

\end{thebibliography}
\end{document}